\documentclass{article}
\usepackage{amsfonts}
\usepackage{graphicx}

\newcommand{\adj}[1]{\mbox{$#1^\dag$}}
\newcommand{\ket}[1]{\mbox{$|#1\rangle$}}

\newcommand{\inner}[2]{\mbox{$\langle#1|#2\rangle$}}

\newcommand{\vect}[1]{\underline #1}

\newtheorem{thm}{Theorem}[section]
\newtheorem{definition}[thm]{Definition}
\begin{document}
\title{Maximum Likelihood Based Quantum Set Separation}%\titlenote{}}
%\titlerunning{ML Based Quantum Set Separation}
%\toctitle{ML Based Quantum Set Separation}
\author{S\'andor Imre, Ferenc Bal\'azs\footnote{The research project was supported by OTKA, id. Nr.: F042590}\\%}
%\authorrunning{S. Imre et al.}
%\institute{
Mobile Communications \& Computing Laboratory
\\Department of Telecommunications\\  Budapest University of
Technology and Economics\\ 1117 Budapest, Magyar Tud\'osok krt.
2, HUNGARY\\
\scriptsize\tt{email: imre@hit.bme.hu,
          balazsf@hit.hit.bme.hu}}
%\\http://www.mcl.hu/\~{}balazsf
%\date{Received: 13 January 1997 / Accepted: 29 September 1997 \\[1em]
%      Communicated by: P. Deuflhard}
%\tocauthor{S\'andor Imre (Budapest University of Technology and
%Economics), Ferenc Bal\'azs (Budapest University of Technology and
%Economics)}
\date{08. December 2003}

%\mainmatter
\maketitle \begin{abstract}In this paper we introduce a method,
which is used for set separation based on quantum computation. In
case of no a-priori knowledge about the source signal
distribution, it is a challenging task to find an optimal decision
rule which could be implemented in the separating algorithm. We
lean on the Maximum Likelihood approach and build a bridge between
this method and quantum counting. The proposed method is also able
to distinguish between disjunct sets and intersection
sets.\end{abstract}
 \section{Introduction} In the course of signal and/or
data processing fast classification of the input data is often
helpful as a preprocessing step for decision preparation. Assuming
that the to be classified data $\mu\in M$ is well defined and it
came under a given number of classes or sets, $A:=\{\mu\in M :
\mathcal{A}(\mu)\}, B:=\{\mu\in M :\mathcal{B}(\mu)\}, \dots,
Z:=\{\mu\in M :\mathcal{Z}(\mu)\}$. To perform the classification
is in such a way equivalent to a set separation task.

The problem of separation could be manifold: sparsely distributed
input data makes the determination of the decision lines between
the classes to a hard (often nonlinear) task, or even the
probability distribution of the input data is not known a-priori
which is resulted in an unsupervised classification problem also
known as clustering \cite{Fuk72}. Further "open question" is to
classify input sequences in the case of only the original
measurement/information data is known almost sure, but the
observed system adds a stochastically changing behavior to it, in
this manner the classification becomes a statistical decision
problem, which could be extremely hard to solve if the number of
"possibilities" is increasing. Due to this fact to find an optimal
solution is time consuming and yields broad ground to suboptimal
ones. With assistance of quantum computation we introduce an
optimal solution whose computational complexity is much lower
contrary to the classical cases.
\par
This paper is organized as follows. In Sect. \ref{sec:comp}. the
set separation related quantum computation basics are highlighted.
The system model is described in Sect. \ref{sec:system}. together
with the proposed set separation algorithm in Sect. \ref{sec:sep}.
The main achievements are revised in Sect. \ref{sec:conc}.
\section{Quantum Computation}\label{sec:comp}
\par In this section we give a brief overview about quantum computation which is relevant to this paper.
For more detailed description, please, refer to
\cite{Sho98,Deu00,Nie00,Eke00}.
\par In the classical information theory the smallest information
conveying unit is the \textit{bit}. The counterpart unit in
quantum information is called the \textit{"quantum bit"}, the
qubit. Its state can be described by means of the state
$\ket{\varphi}$, $\ket{\varphi}=\alpha\ket{0}+\beta\ket{1}$, where
$\alpha,\beta \in \mathbb{C}$ refers to the complex probability
amplitudes and $|\alpha|^2+|\beta|^2=1$ \cite{Sho98,Deu00}. The
expression $|\alpha|^2$ denotes the probability that after
measuring the qubit it can be found in  computational base
$\ket{0}$, and $|\beta|^2$ shows the probability to be in
computational base $\ket{1}$. In more general description an
$N$-bit \textit{"quantum register"} (qregister) $\ket{\varphi}$ is
set up from qubits spanned by $\ket{x}$ $x=0\dots(N-1)$
computational bases, where $N=2^n$ states can be stored in the
qregisters at the same time \cite{imr01}
\begin{equation}
\ket{\varphi}=\sum_{x=0}^{N-1}\varphi_x\ket{x}; ~~~\varphi_x\in
\mathbb{C}, \label{eq:1}\end{equation} where $N$ denotes the
number of states and $\forall x\neq j$, $\inner{x}{j}=0$,
$\inner{x}{x}=1$, $\sum|\varphi_x|^2=1$, respectively. It is worth
mentioning, that a  transformation $U$ on a qregister is executed
parallel on all $N$ stored states, which is called \textit{quantum
parallelizm}. To provide irreversibility of transformation, $U$
must be unitary $U^{-1}=\adj{U}$, where the superscript $(\dag)$
refers to the Hermitian conjugate or adjoint of $U$. The quantum
registers can be set in a general state using quantum gates
\cite{Nie00,Eke00} which can be represented by means of a unitary
operation, described by a quadratic matrix.

%\begin{figure}[tb]
%\begin{center}
%\includegraphics[width=50mm]{iter1.eps}
%\includegraphics[width=90mm, height=35mm]{counting.eps}
%\end{center}
%\caption{The quantum counter circuit}\label{fig:counter}
%\end{figure}

\section{System Model}\label{sec:system}
For the sake of simplicity a 2-dimensional set separation is
assumed, where the original source data can take the values $\mu
\in \mathbf{M}^{[0,1]}$ and was chosen from the sets $s=0$ and
$s=1$. Additional information on the source is not available, e.g.
also nothing about the probability density function
(\textit{pfd}).
%\begin{defi}
\begin{figure}[tb]
\begin{center}
\includegraphics[width=65mm, height=30mm]{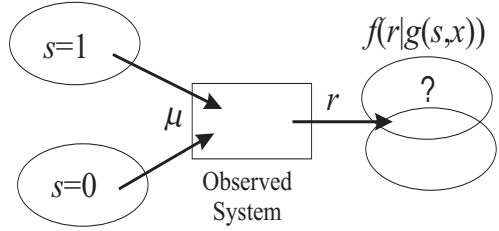}
\end{center}
\caption{General set separation system}\label{fig:syst}
\end{figure}
The general set separation system is depicted in Fig.
\ref{fig:syst}. The observed signal $r$, disturbed by the system
$A$, becomes the input data which will be separated into the two
sets ($s=0$ and $s=1$) again.
\par
\begin{figure}[tb]
\begin{center}
\includegraphics[width=65mm, height=25mm]{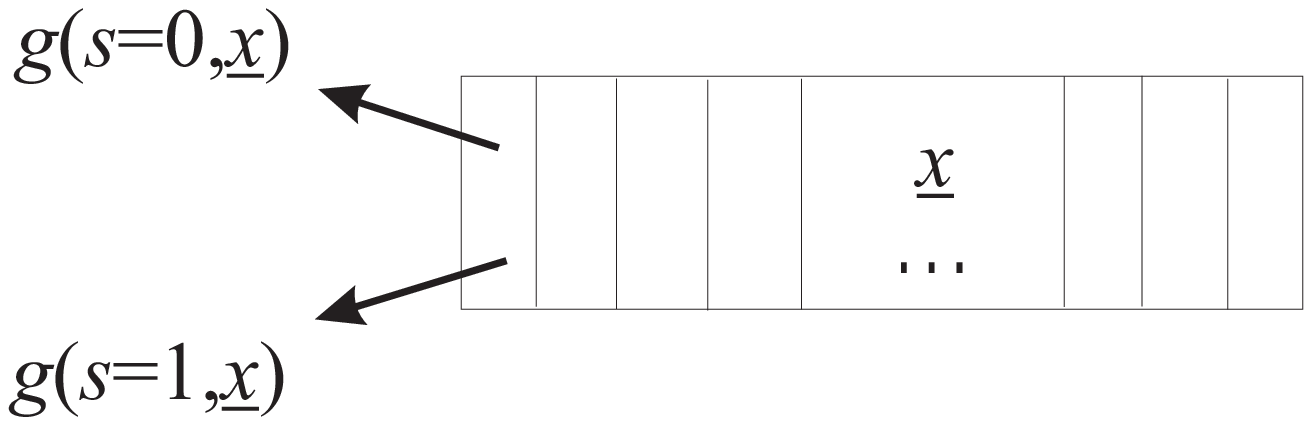}
\end{center}
\caption{Quantum register $\ket{\varphi}$}\label{fig:qreg}
\end{figure}

 In the set separator a quantum register $\ket{\varphi}$ --as
described by equation (\ref{eq:1}) and shown in Fig.
\ref{fig:qreg}.-- is used to store all the parameters, e.g. delay,
heat, velocity, etc. values of the possible system disturbance in
a specially given quantization\footnote{Quantization is NOT a
quantum computation operation!}\footnote{The quantization method,
i.e. linear or nonlinear is out of the scope of this paper.}. As
an example: in the qregister $\ket{\varphi}$, the properly
prepared, quantized delay and velocity values are stored, e.g. the
values $1.0\cdot 10^{-1}, 1.1\cdot 10^{-1},\ldots, 1.0\cdot
10^{-10}$ and $1.0 ~m/s, 1.1 ~m/s, \ldots, 100 ~m/s$. This
information is not utilizable so far but the combination of this
effects, i.e. this values, whose extent could blast any database.
To handle the large amount of data to be processed a virtual
database should be introduced. \vspace*{3mm}
\begin{definition}
To build up a virtual database a function
\begin{equation}
y=g(s,\vect{x}), \label{eq:2}
\end{equation}
is defined, where $s\in S$ identifies the sets and $\vect{x}$
denotes the index of the qregister $\ket{\varphi}$, respectively.
The function $y_i=g(s,x_i)$ points to an record in the virtual
database.
\end{definition}
%\vspace*{2mm}
%\paragraph{
\subsection{
Properties of the Function $g(\cdot)$}\label{ssec:prop} The
function $g(s,\vect{x})$ is not obligingly mutual unambiguous
consequently, it is not reversible, except for several special
cases, when the virtual database contains
$\widehat{r}=g(s,\vect{x})$ only once. In this case the parameter
settings of the system $A$ are easy to determine. Nevertheless,
the fact to have an entry only once in the virtual database
described by the equation $g(s_i,\vect{x})$ does not exclude to
have the same entry in other virtual databases generated by
$g(s_j,\vect{x})$, where $i\neq j$, which makes a trivial decision
impossible. Henceforth the fact should be kept in mind that
$g(s,\vect{x})$ is in almost every case a so called one way
function which is easy to evaluate in one direction, but to
estimate the inverse is rather hard. \par The function $g(.)$
generates all the possible disturbances additional to the
considered input value $\mu$ belonging to the set $s=0$ or $s=1$
of the system. This is of course a large amount of information,
$2N=2^{n+1}$, where $n$ is the length of the qregister
$\ket{\varphi}$. For an example let us assume a 15-qbit qregister.
The function $g(\cdot)$ in (\ref{eq:2}) generates $2^{15}=32.768$
output values at the same time for $s=0$ and the same number of
outputs for $s=1$. Taking into account the large number of
possible points in the set surface the optimal classification in a
classical way becomes difficult.

%It is not necessary to assign
%$\widehat{\mu}=r$ to the most likely set $S$, but all the possible
%states should be evaluated using the function (\ref{eq:2}) and
%have to be compared to the input value disturbed by the observed
%system.
At the first glace this problem looks more difficult to solve,
however, with exploiting the enormous computational power of
quantum computation, in this case the Deutsch-Jozsa \cite{Deu92}
quantum parallelization algorithm, an arbitrary unitary operation
can be executed on all the prepared states contemporaneously.

\subsection{Quantum Search in Qregister
 $ \ket{\varphi}$} \label{ssec:qsearch}
\begin{figure}[b]
\begin{center}
\includegraphics[width=50mm]{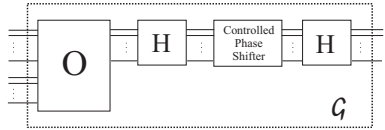}
\end{center}
\caption{The Grover database search circuit}\label{fig:grover1}
\end{figure}

Roughly speaking the task is to find the entry (entries) in the
virtual databases which is (are) equal to the observed data $r$.
To accomplish the database search the Grover database search
algorithm should be invoked \cite{Gro96}. In Sect.~\ref{sec:comp}.
we proposed to set up an qregister, which has to be built up only
one time at all. It is obvious to choose a suitable database
searching algorithm, to see which function
$g_{0,1}(s_{0,1},\vect{x})$ picking the vector $\vect{x}$ form
qregister $\ket{\varphi}$ contains the searched \textit{bit}, if
any at all. We apply the optimal quantum search algorithm
$\mathcal{G}$, as depicted in Fig.~\ref{fig:grover1}. proposed by Grover \cite{Gro99,Zal99}. %and depicted in
%Figure \ref{fig:grover} for our purpose.
We feed the received signal $r(t)$ %(\ref{eq:mud1})
to the oracle $(\mathrm{O})$, where the function
$f(r,g(s,\vect{x})$$)$ is evaluated such that
\begin{equation}
f(a,b)= \left\{
\begin{array}{cl}
\mbox{1} & \mbox{if $a=b$}
\\ 0 & \mbox{otherwise}.
\end{array}\right. \label{eq:20}
\end{equation}
%where $r^k=\sqrt{E_k}a_k\int_0^{T_s}r(t)s_k(t)\mathrm{d}$$t$.
Assuming, there is again $M$ solutions for the search in qregister
$\ket{\varphi}$,
\begin{equation}
\ket{\varphi}=\sqrt{\frac{N_{s}-M}{N_{s}}}\ket{\alpha}+\sqrt{\frac{M}{N_{s}}}\ket{\beta},
\end{equation}
where $\ket{\alpha}$ consists of such configurations of $\ket{x}$,
 which does not results $\widehat{\mu}=r$, while
$\ket{\beta}$ does.

%Since, the number of solutions $M$ is unknown, however, $M\ll
%N_{s}$, presumably, \cite{boy96} gave a tight upper bound for the
%number of iterations, that is $l\approx
%\frac{9}{2}m_0=\frac{9}{2}\cdot \frac{1}{\sin
%\varphi}<\frac{9}{2}\sqrt{\frac{N_s}{M}}$, for
%$0<M<\frac{3N_s}{4}$, which is still
%$\mathcal{O}(\sqrt{\mathrm{N}})$.

Because of the fact of tight bound, in real application less
iterations would be also appropriate \cite{Imr03}.

\section{Set Separation}\label{sec:sep}
Let us turn our interest back to the separation of the observed
data $r$ from the predefined sets.

Assuming the special case where only one of the virtual database
descriptor functions, either $g(s_{0},\vect{x})$ or
$g(s_{1},\vect{x})$ contains the entry identical to the observed
data $r$ a set separation can be performed easily. \par
\begin{figure}[tb]
\begin{center}
\includegraphics[width=25mm, height=20mm]{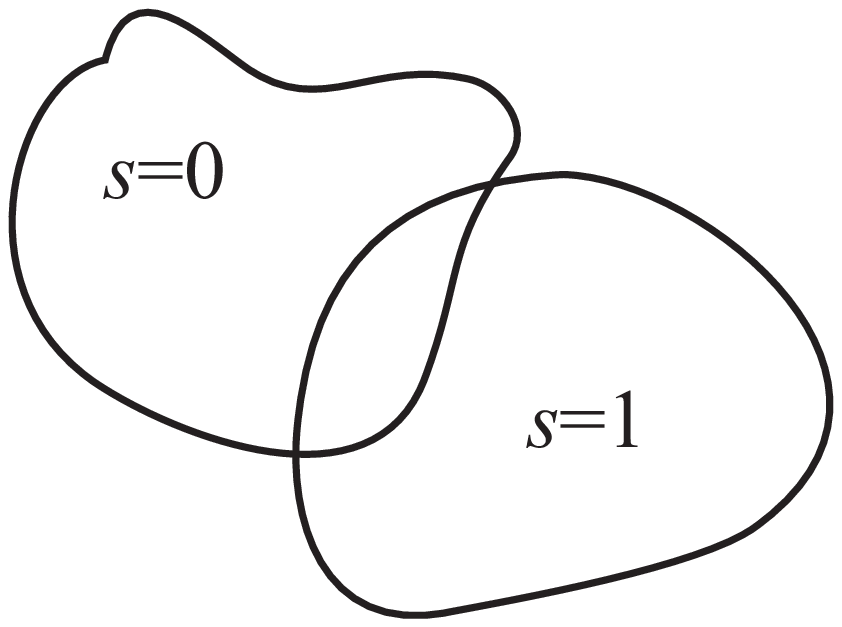}
\end{center}
\caption{Sets with intersection}\label{fig:set2}
\end{figure}

 A more
realistic case is to have an intersection part of the two sets as
shown in Fig.~\ref{fig:set2}.  Even so, due to passing the
observed system, overlapping of the sets can be occurred due to
disturbances. After evaluating the functions
$g_{0,1}(s_{0,1},\vect{x})$ it could happen that the same records
are multiple present, which shows the irreversibility behavior of
the function (\ref{eq:2}). Originally, the input signal was chosen
from well defined disjunct sets without a-priori known probability
distributions. The process, to put $r$ to a set either to $s_0$ or
to $s_1$ should be based on \textit{Maximum Likelihood} decision.

\par Let us assume that we have a random variable $r$. Its
measured value depends on a selected element $x_l$ from a finite
set ($l=1,\ldots,L$) and a process which can be characterized by
means of a conditional pdf $f(r|x_l)$ belonging to the given
element. Our task is to decide which $x_l$ was selected if a
certain $r$ has been measured. Each guess $H_l$ for $x_l$ can be
regarded as a hypothesis. Therefore  decision theory is dealing
with design and analysis of suitable rules building connections
between the set of observations and hypotheses.

\par If we are familiar with the unconditional  (a priori) probabilities $P(x_l)$ then
the Bayes formula helps us to compute the conditional (a
posteriori) probabilities $P(H_i|r)$ in the following way
\[P(H_l|r)=\frac{f(r|x_l)P(x_l)}{\sum^L_{i=1}f(r|x_i)P(x_i)}.\]
Obviously the most pragmatic solution if one chooses $H_l$
belonging to the largest $P(H_l|r)$. This type of hypothesis
testing is called \emph{maximum a posteriori} (MAP) decision.

\par If the a priori probabilities are unknown or $x_l$ is equiprobable then \emph{maximum likelihood}
(ML) decision can be used. It selects $H_l$ resulting the largest
$f(r|x_l)$ when the observed $r$ is substituted in order to
minimize the probability of error

%\par The following theorem is exploited when we are designing optimal multiuser detectors in
%Chapter \ref{ch:qmud}
%
\[\max_{l} L(r,x_l).\]
The Maximum Likelihood estimator requires to know the probability
density function of the observed signal. Employing the Grover
database search algorithm we are able to find the entries in the
virtual databases, however, it is not needed to perform a complete
search because the search result --the exact index (indices) of
the searched item(s)-- is (are) not interesting but the number how
often a given configuration is involved in $g(s,\vect{x})$ or not.
For that purpose a new function $f(\cdot)$ is defined.
\vspace*{3mm}
\begin{definition}
The function
\begin{equation}
f(r|s)=\frac{\sharp\left(x: r=g(s,x)\right)}{\sharp (x)},
\label{eq:9}
\end{equation}
 counts the number of similar entries in the virtual
database, which corresponds to the conditional probability density
function $r$ to be in the set $s$.
\end{definition}
\vspace*{2mm} For that reason it is worth stepping forward to
quantum counting \cite{Bra98_2} based on Grover iteration.

\subsection{Set Separation Method}
\begin{figure}[tb]
\begin{center}
\includegraphics[width=85mm, height=40mm]{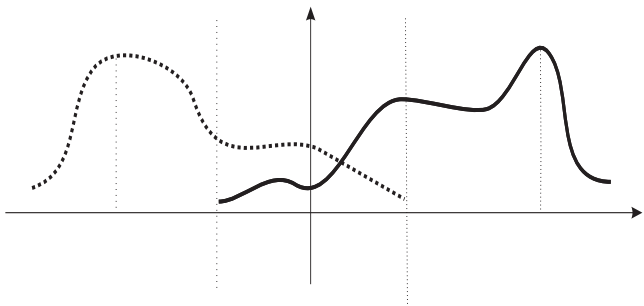}
\end{center}
\caption{The two density functions $f\left(r|s=0\right)$ and
$f\left(r|s=1\right)$ }\label{fig:inter2}
\end{figure}

\par The both curves in Fig.~\ref{fig:inter2}. represent the number
of the same entries in the virtual databases, i.e. the pdf's,
according to $f(r|s=0)$ and $f(r|s=1)$, respectively. In case of
having entry(entries) only in $y_i$ but not in $y_j$ of function
$g_{0,1}(s_{0,1},\vect{x})$, where $i,j\in [0,1]$, and $i\neq j$,
means a 100 percent sure decision, following the decision rules in
Table~\ref{tab:1}. This areas are the non-overlapping parts of the
sets in Fig.~\ref{fig:set2}. and the outer parts (until the
vertical dashed black lines) in Fig.~\ref{fig:inter2}. However, in
the case of non zero $f(r|s=0)$ and $f(r|s=1)$ values an accurate
prediction can be given relating to the Maximum Likelihood
decision rule.

\begin{table}[h]%[tb]
\begin{center}
\caption{Set Separation Decision Rules} \label{tab:1}
\begin{tabular}[t]{c|c|c} \hline
$f(r|s_0)$ & $f(r|s_1)$ & Decision
\\\hline 0 & 0 & $\ket{\varphi}$ was badly prepared \\ 0 & $\neq 0$ & $r$ belongs to set $s=1$\\
$\neq 0 $ & 0 & $r$ belongs to set $s=0$ \\ \cline{1-2}
\multicolumn{2}{c|}{$>$} & $r$ belongs to set $s=0$ \\
\multicolumn{2}{c|}{$<$}  & $r$ belongs to set $s=1$ \\ \hline
\end{tabular}
\end{center}
\end{table}
\par All the possible states from the qregister $\ket{\varphi}$
will be evaluated by the function (\ref{eq:2}) for $s=0$ and also
for $s=1$, simultaneously, which will be collated with the system
output $r$. If at least one output $y_0$ or $y_1$ with the
parameter settings $x$ is matched to the system output $r$, it
will be put to the set $s=0$ or $s=1$, respectively. In a more
exciting case at least one similarity of $y_0$ and also at least
one of $y_1$ to $r$ is given, the system output could be
classified to the both sets, an intersection is drawn up. This
result in a not certainty prediction, which piques our interest
and sets our focus not this juncture.

We assume no a-priori knowledge on the probability distribution of
the input sequence $\mu$, so it is assumed to be equally
distributed. Henceforward we suppose that after counting the
evaluated values $f(r|s=i)$ the number of similarity to the system
output $r$ is higher than in case of $f(r|s=j)$, where $i,j\in
[0,1]$. In pursuance of the decision rule in Table~\ref{tab:1}.,
$r$ belongs rather to set $s=i$ than to set $s=j$.

\paragraph{The Method}
To perform a set separation nothing else is required as
\begin{enumerate}
\item Prepare the qregister $\ket{\varphi}$, \item Evaluate the
functions $y_i=g_i=(s=i,\vect{x})$, where $i\in [0,1]$ in
2-dimensional case, \item Count the identical entries in the
virtual databases which are equal to the observed data $r$,
$f(r|s)$, (see Fig. \ref{fig:inter2})., \item Use the decision
table Table \ref{tab:1} to assign $r$ to the sets $s=0$ or $s=1$.
\end{enumerate}

\section{Concluding Remarks}\label{sec:conc}
In this paper we showed a connection between \textit{Maximum
Likelihood} hypothesis testing and Quantum Counting used for
quantum set separation. We introduced a set separation algorithm
based on quantum counting which was employed to estimate the
conditional probability density function of the observed data in
consideration to the belonging sets. In our case the
\textit{pdf's} are estimated fully at a single point by invoking
the quantum counting operation only once, that makes the decision
facile and sure. In addition one should keep in mind that the
qregister $\ket{\varphi}$ have to be set up only once before the
separation. The virtual databases are generated once and directly
leaded to the Oracle of the Grover block in the quantum counting
circuite, which reduce the computational complexity,
substantially.

%\bibliographystyle{ieeetran}
%\bibliography{qubit,bib1}
\end{document}